\documentclass[twocolumn,prl,tightenlines,superscriptaddress,showpacs]{revtex4-1}

\usepackage{amsmath}
\usepackage{amssymb,amsfonts,latexsym}
\usepackage{bm}
\usepackage[mathcal]{euscript}
\usepackage{graphicx}
\usepackage{epsfig}
\usepackage{color}
\usepackage{xfrac}
\usepackage{mathrsfs}

\usepackage{hyperref}
\usepackage{url}

\begin{document}

\title{Small Obstacle in a Large Polar Flock}

\author{Joan Codina}
\affiliation{Wenzhou Institute, University of Chinese Academy of Sciences, Wenzhou 325001, China}
\affiliation{Key Laboratory of Soft Matter Physics, Institute of Physics, Chinese Academy of Sciences, Beijing 100190, China}

\author{Beno\^{\i}t Mahault}
\affiliation{Max Planck Institute for Dynamics and Self-Organization (MPIDS), 37077 G\"ottingen, Germany}

\author{Hugues Chat\'{e}}
\affiliation{Service de Physique de l'Etat Condens\'e, CEA, CNRS Universit\'e Paris-Saclay, CEA-Saclay, 91191 Gif-sur-Yvette, France}
\affiliation{Computational Science Research Center, Beijing 100193, China}
\email{hugues.chate@cea.fr}

\author{Jure~Dobnikar}
\affiliation{Key Laboratory of Soft Matter Physics, Institute of Physics, Chinese Academy of Sciences, Beijing 100190, China}
\affiliation{School of Physical Sciences, University of Chinese Academy of Sciences, Beijing 100049, China}
\affiliation{Songshan Lake Materials Laboratory, Dongguan, Guangdong 523808, China}
\email{jd489@cam.ac.uk}

\author{Ignacio Pagonabarraga}
\affiliation{Departament de F\'{\i}sica de la Mat\`eria Condensada, Universitat de Barcelona, 08028 Barcelona, Spain}
\affiliation{Universitat de Barcelona Institute of Complex Systems,
%Universitat de Barcelona, 
08028 Barcelona, Spain}
\affiliation{Centre Europ\'{e}en de Calcul Atomique et Mol\'{e}culaire, Ecole Polytechnique F\'{e}d\'{e}rale de Lausanne, 1015 Lausanne, Switzerland}
\email{ipagonabarraga@ub.edu}

\author{Xia-qing Shi}
\affiliation{Center for Soft Condensed Matter Physics and Interdisciplinary Research, Soochow University, Suzhou 215006, China}
\email{sxiaqing@foxmail.com}

\date{\today}

\begin{abstract}
We show that arbitrarily large polar flocks are susceptible to the presence of a
single small obstacle. In a wide region of parameter space, the obstacle  
triggers counter-propagating dense bands leading to reversals of the flow.
In very large systems, these bands interact yielding a never-ending chaotic dynamics that
constitutes a new disordered phase of the system.
While most of these results were obtained using simulations of aligning self-propelled particles, we find similar phenomena at the continuous level, not when considering the basic Toner-Tu hydrodynamic theory, but in simulations of truncations of the relevant Boltzmann equation.
\end{abstract}

\maketitle
 
In numerical and theoretical studies of active matter, ‘polar flocks’ continue to play a central role
(see, {\it e.g.} \cite{Souslov_van_Zuiden_Bartolo_Vitelli_2017,gomez2018collective,kourbane2018exact,geyer2019freezing,mahault2019TT,sone2019anomalous,dadhichi2020nonmutual,tasaki2020hohenberg,martin2021fluctuation,martin2021active_vortices,sansa2021phase_separation,zhao2021phases,yu2021breakdown} 
for recent examples). 
This term refers to the homogeneous collective motion,
resulting from spontaneous rotational symmetry breaking,
 of self-propelled particles 
locally aligning their velocities, as in the Vicsek model \cite{vicsek1995novel,ginelli2016physics,chate2020dry}.
Remarkably, polar flocks exhibit true long-range polar order even in two space dimensions (2D), 
as argued by Toner and Tu, who also predicted the scaling structure of their space-time fluctuations
\cite{toner1995long,toner1998flocks,toner2012reanalysis,mahault2019TT}. 
Since these early works, a wealth of results have been obtained on particle-level models, hydrodynamic theories,
and even experimental realizations of polar flocks
\cite{deseigne2010collective,weber2013long,kumar2013flocking,soni2020phases,geyer2018sounds,iwasawa2021algebraic}. 
%demonstrating robustness and genericity of their properties. 
The overall phase diagram of such dry aligning dilute active matter is best described as
resulting from a phase-separation scenario in which homogeneous polar flocks form an 
orientationally-ordered liquid separated from a disordered gas by a coexistence phase involving dense and ordered travelings bands \cite{solon2015phase,DADAM_LesHouches,chate2020dry}. 
This has been recently shown to hold even in the case of non-metric, `topological' interactions, 
a case that was heretofore believed to show a direct order-disorder transition \cite{martin2021fluctuation}.

Polar flocks are thus observed in a rather large class of active matter systems and there is evidence of the 
robustness of their properties.
 But there are also clear indications of their fragility. For instance even weakly non-reciprocal interactions
have been shown to substantially modify phase diagrams \cite{chen2017fore,dadhichi2020nonmutual,fruchart2021non-reciprocal}.
Further evidence 
%of fragility of polar flocks 
is found in recent works that demonstrated
a variety of effects induced by spatial quenched disorder
\cite{chepizhko2013optimal,toner2018swarming,toner2018hydrodynamic,yu2021breakdown,chardac2021emergence}.
The Toner-Tu long-range polar order has been argued to be qualitatively changed for flocks moving on disordered substrates. It has been shown that different types of quenched disorder can have different effects. 
In most cases, arbitrarily weak disorder modifies or breaks long-range polar order. 

The above cases of quenched disorder include that of scattered small obstacles, 
which was considered in models and in experiments
\cite{chepizhko2013optimal,chardac2021emergence}.
But what about the effect of a single obstacle? 
The insertion of passive objects in a disordered gas of active particles is known to have 
non-trivial consequences, introducing in particular long-range currents in the active gas
\cite{baek2018generic,granek2020bodies,rohwer2020activated,knezevic2020effective,liu2020constraint,sebtosheikh2020effective,speck2021vorticity}.
Given this, one can expect that even a single passive obstacle introduced within a long-range correlated polar flock 
may have major consequences on the large-scale dynamics of the system.

%%%%%%%%%%%%%%%%%%%%%%%%%%%%%%%%%%%%%%%%%%%%%%%%%%%%
\begin{figure*}[t!]
\includegraphics[width=\textwidth]{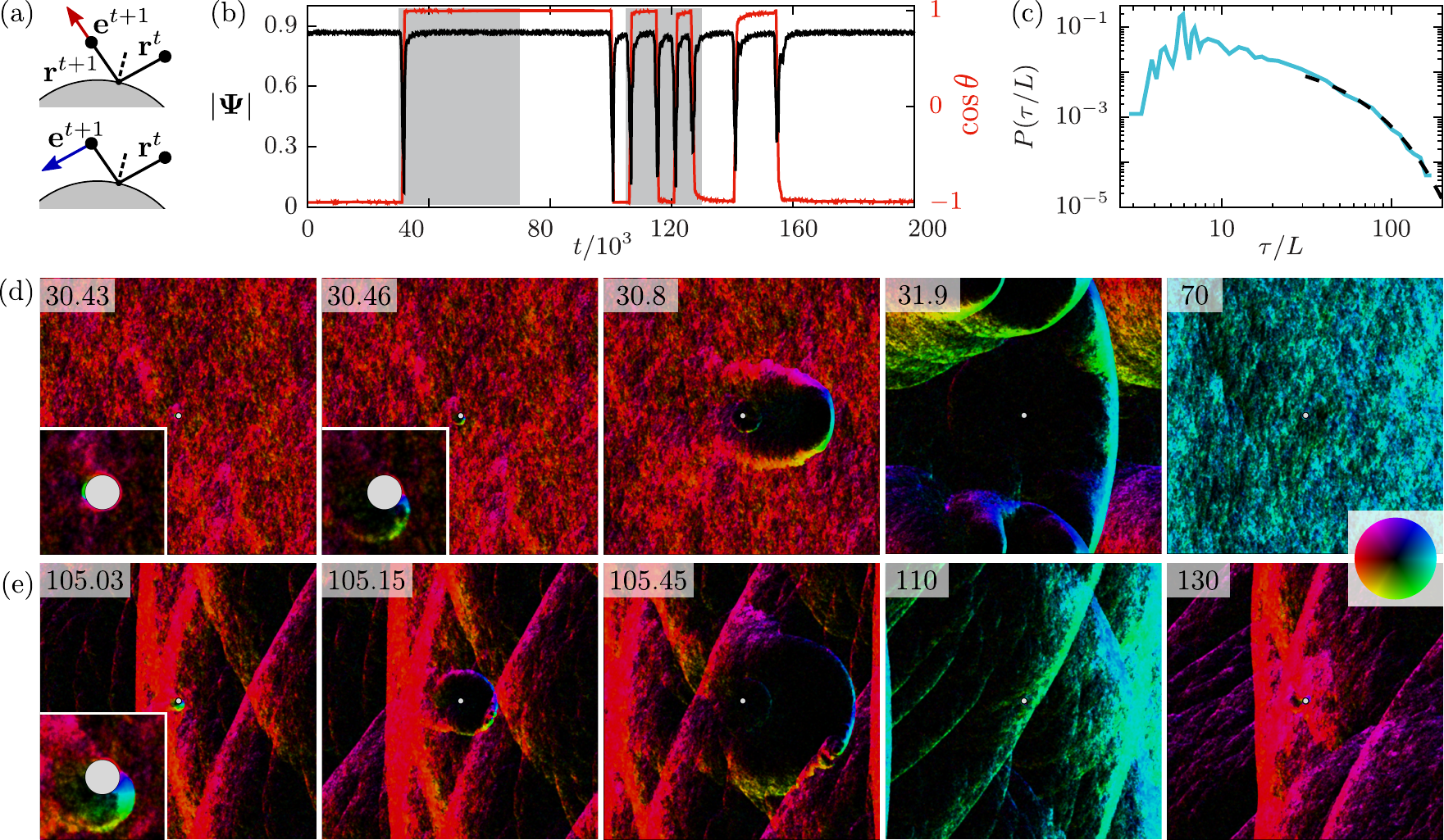}
\caption{Typical flow reversals in a system of size $L=1024$ at $\eta=0.5$ with $\sigma=22$.
(a) Sketch of collision rules:
Orientation is either reflected (type 1, in red) or kept (type 2, in blue).
(b) Time series of the global polarization $|\bm \Psi|$, in black, and its orientation $\cos\theta$, in red. Shaded regions indicate the time windows for the snapshots series in (d), and (e).
(c) Distribution of Inter-reversal times $P(\tau)$. Dashed line indicates an exponential decay.
(d) Snapshots of the system during an isolated reversal, and its recovery. 
On the color wheel on the right, color and intensity represent, respectively, the angle of the locally averaged polarity
and the locally averaged density (with black for near-empty regions)
The obstacle (at the center of the images) is in grey. Time marks are indicated at the top of each panel in units of 1000 timesteps. 
For the first 2 panels, a zoom is provided showing the birth of the counter-propagating front around the obstacle.
(e) Same as (d) but during a sequence of reversals triggered by the passage of the dense band over the obstacle.
See Movies~S1 and S2 in \cite{SUPP} related to (d) and (e).
}
\label{fig1}
\end{figure*}
%%%%%%%%%%%%%%%%%%%%%%%%%%%%%%%%%%%%%%%%%%%%%%%%%%%%

In this Letter, we explore the consequences of the introduction of a single fixed obstacle in the flow exhibited by 2D polar flocks using both self-propelled particles models and continuous theories. We show that, in a large
part of the ordered liquid phase, even a small disk can reverse the global orientation of very large flocks.
Such reversals occur via the nucleation, ballistic invasion, and complex interactions of counter-propagating dense bands. 
For very large systems, reversals become complex events of increasing duration, 
and we conjecture that asymptotically the flock never repairs itself.
Our data indicate that orientational order is fully broken and self-averages in the infinite-size limit, albeit with very large correlation scales.  
At the continuous level, we find that the standard Toner-Tu hydrodynamic theory is unable
to account for these phenomena, something low-order truncations of the Boltzmann equation for polar flocks can do.

In most of this work, we make use of a standard 2D Vicsek model with vectorial noise \cite{gregoire2004onset}: 
point particles $i=1,2,\ldots,N$, with positions ${\bf r}_i$ and unit orientations ${\bf e}_i$ 
move at constant speed $v_0$ in discrete time steps:
\begin{subequations}
\begin{eqnarray}
	{\bf r}_i^{t+1} &=& {\bf r}_i^t + v_0 {\bf e}_i^{t +1}, \label{eqvm1}\\
	{\bf e}_i^{t+1} &=& \vartheta\left[
	\langle {\bf e}_j^t\rangle_{j\sim i} + \eta \, {\bm \xi}
	\right],
	\label{eqvm2}
\end{eqnarray}
\label{eqvm}
\end{subequations}
where $\vartheta$ normalizes vectors ($\vartheta({\bf u})={\bf u}/\|{\bf u}\|$), 
the average is taken over all particles $j$ within unit distance of $i$ including $i$,
and ${\bm \xi}$ is a random-orientation unit vector drawn independently for each particle at each timestep.
This may seem an odd choice for modeling collisions with an obstacle since particles have no physical size, and discrete-time dynamics look inconvenient. 
Our motivation here, though, is to be able to reach the asymptotic regime of polar flocks, 
which is known to require very large system sizes, even for Vicsek-like systems \cite{mahault2019TT}, and remains
inaccessible with more realistic particles.

In this Vicsek context, collisions with an obstacle can be implemented in various ways. 
We have considered two types of interactions (cf. Fig.~\ref{fig1}(a), with more details in \cite{SUPP}). 
Colliding particles are those whose next position is calculated to land within the obstacle. 
This location is replaced by that given by a simple reflection on the obstacle. The two cases we considered
are distinguished by the choice of ${\bf e}_i^{t+1}$, the orientation of the particle after collision with the obstacle. 
In Type~1 collisions, this orientation is simply given by the reflected trajectory 
used to calculate ${\bf r}_i^{t+1}$.
In Type~2 collisions, the initially calculated orientation is kept, as for circular active particles
(endowed with an intrinsic polarity axis) bumping on a hard surface (see e.g. \cite{weber2013long}).
Despite these collision rules being rather different, we found that they do not lead to any significant difference in the behavior of the system. 
Below we use Type 2 collisions, while some results obtained with Type 1 collisions are shown in \cite{SUPP}.

We consider a single fixed circular obstacle of diameter $\sigma$ 
in a square domain of linear size $L$ with periodic boundary conditions.
For simplicity, we set the global density $\rho_0=N/L^2=2$, $v_0=1$,
and vary the remaining important parameter, 
the noise strength $\eta$. The pure, obstacle-less system, exhibits the 3 expected phases as $\eta$ decreases: 
a disordered gas for $\eta\gtrsim 0.65$, the homogeneous Toner-Tu polar flock phase for $\eta\lesssim 0.56$, and the coexistence phase with its signature traveling bands in between. 
Here, we are mostly concerned with the polar flock phase. 

Monitoring the global order 
parameter ${\bm \Psi}=\langle {\bf e}_i\rangle_i$, 
%which is typically oriented along one direction of the box,
we observe that any obstacle of diameter $\sigma$ 
significantly larger than the unit interaction length 
\footnote{We observed reversals with obstacles as small as about ten times the unit interaction length. This minimal obstacle size, though, likely depends on the model and collision rules.}
triggers sudden changes in $\theta=\arg{\bm \Psi}$, the global flow direction, accompanied
by sharp dips of $|{\bm \Psi}|$ (Fig.~\ref{fig1}(b)). 
Most of these events are full reversals during which $\theta$ changes by $\pi$, 
but some result in $\tfrac{\pi}{2}$ rotations. For simplicity, we call all of them reversals hereafter.

We observe two types of reversals. 
In the regime shown in Fig.~\ref{fig1}, the flow around the obstacle recovers a 
homogeneous steady state between reversals that are sufficiently far apart in time. 
Our first type of reversal are those nucleated from this steady state, roughly as follows: 
An initial dense, counter-propagating blob of particles is first nucleated near the obstacle surface, at its rear. 
It then recruits more and more particles as it moves along and detaches from the obstacle, 
forming a dense, curved band which invades
the whole system ballistically. In the final stage, the global polarity is now typically reversed, and this band, which has now connected itself across the periodic boundaries, widens slowly (Fig.~\ref{fig1}(d), and Movie~S1 in \cite{SUPP}). 
This slow recovery can proceed until the homogeneous steady state is recovered, but it can also be interrupted 
at the occasion of one of the multiple passages of the band `through' the obstacle. 
As shown in Fig.~\ref{fig1}(e)  and Movie~S2 in \cite{SUPP}, a passing band can trigger a new counter-propagating dense front which can reverse (again) the global polarity. This constitutes our second type of reversal.

\begin{figure}[t!]
\includegraphics[width=\columnwidth]{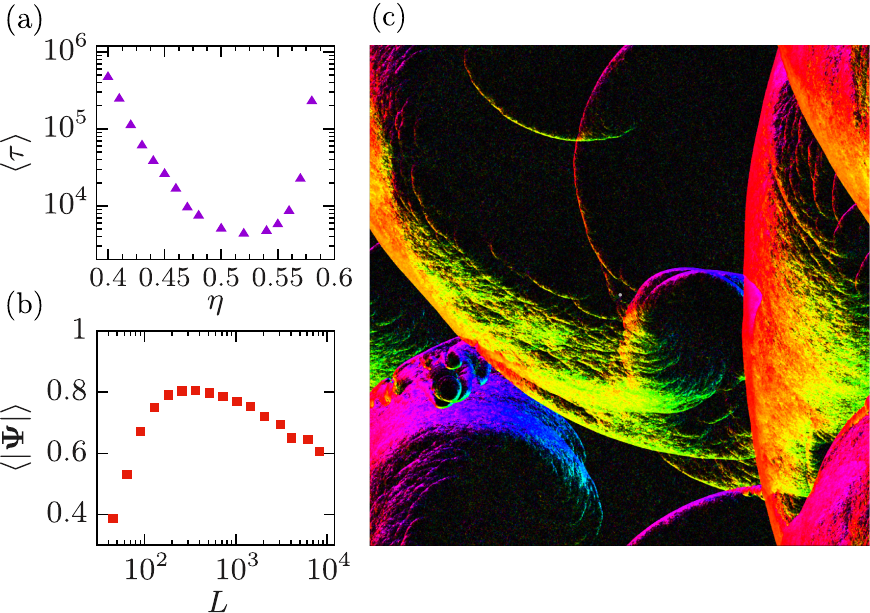}
\caption{
(a) Average inter-reversal time, $\langle\tau\rangle$ vs $\eta$ ($L=256$, $\sigma=35$). 
(b) Average global polarization vs system size with obstacle (red squares) and without (dashed line) 
($\eta=0.5$, $\sigma=35$).
(c) Snapshot of a $L=4096$ system ($\eta=0.5$, $\sigma=35$, colors as in Fig.~\ref{fig1}(a)); 
at these parameters the system almost never recovers the homogeneous polar state. 
See related Movie~S3 in \cite{SUPP}
}
\label{fig2}
\end{figure}

For the parameters and system size used in Fig.~\ref{fig1}, second-type reversals are relatively rare,
and the tail of the distribution of inter-reversal times $\tau$ is exponential, 
being dominated by the first-type nucleation events (Fig.~\ref{fig1}(c)). 
This is rather typical, but the relative frequency of the two types of reversals varies with parameters and system size
in a complicated manner, and there are regimes displaying almost periodic second type reversals
\footnote{This complex situation will be detailed in a forthcoming publication.}.
In all cases, though, we observe the same behavior for the variation of $\langle\tau\rangle$ 
with the noise strength $\eta$: 
$\langle\tau\rangle$ takes a minimal value near $\eta\simeq0.5$ and
increases very fast both when decreasing $\eta$ from this value and when approaching 
the boundary of the coexistence band phase (Fig.~\ref{fig2}(a), note the log scale).
The existence of a minimum of $\langle\tau\rangle(\eta)$ can be rationalized as follows: 
decreasing $\eta$, the homogeneous flock phase 
(and the incoming bands, in the case of the second type of reversal)
 is more and more stable, making it harder 
for the nucleated counter-propagating dense group to win.
On the other side, approaching the coexistence phase, there are stronger fluctuations in the incoming flow, 
which lowers the probability of forming an initial counter-propagating group.

All results shown so far have been obtained with system sizes $L$ much larger than the obstacle diameter $\sigma$,
and it is thus clear that a small object can frequently reverse the global order of even a large flock. 
But in cases such as that presented in Fig.~\ref{fig1}, the time-averaged order parameter
$\langle|{\bm \Psi}|\rangle_t$ keeps a fairly large value in spite of reversals, 
essentially because they remain rather well-separated events. 
However, simulating much larger systems reveals that the ballistic expansion of the counter-propagating 
band typically does not lead to simple `reconnection' across the boundaries as it does in Fig.~\ref{fig1}(d). 
Instead, the system can engage in a long, complicated process during which bands cross, or meet the obstacle, 
leading to the nucleation of more bands (Fig.~\ref{fig2}(c), and Movie~S3 in \cite{SUPP}). 
One cannot distinguish reversals anymore and the global dynamics becomes chaotic
as the system size is increased. 
Consequently, $\langle|{\bm \Psi}|\rangle_t$ decreases with increasing $L$, approaching
the scaling of a self-averaging disordered phase (Fig.~\ref{fig2}(b)).
Extrapolating these results, we conclude that a single obstacle asymptotically destroys order.
Data such as those in Fig.~\ref{fig2}(b) could only be obtained for 
`favorable' $\eta$ values near $0.5$, where, at moderate system size, $\langle\tau\rangle$ is not too large.
We nevertheless believe global order is asymptotically destroyed over the larger range 
of noise strengths where we are able to observe reversals. 
This range, however, remains fairly limited.

%%%%%%%%%%%%%%%%%%%%%%%%%%%%%%%%%%%%%%%%%%%%%%%%%%%%
%%%%%%%%%%%%%%%%%%%%%%%%%%%%%%%%%%%%%%%%%%%%%%%%%%%%

\begin{figure}[t!]
\includegraphics[width=\columnwidth]{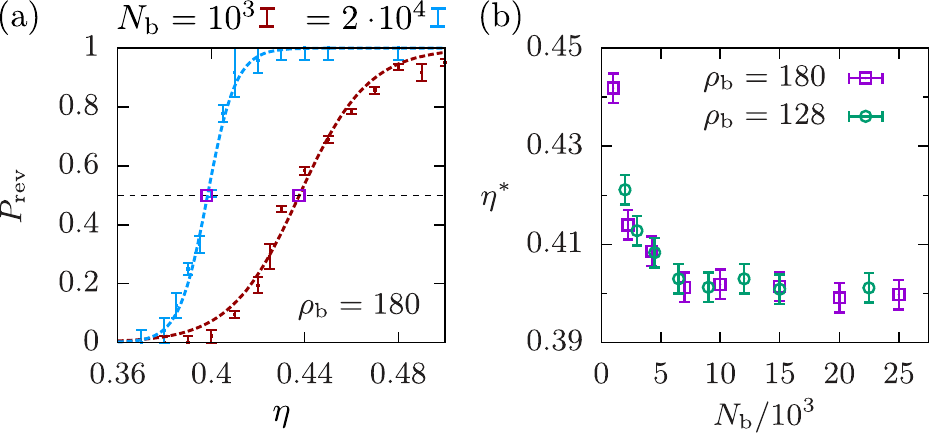}
\caption{Reversals triggered by the introduction of a finite blob of counter-propagating particles ($L=1024$).
(a) Reversal probability $P_{\rm rev}$ vs $\eta$ at different $N_{\rm b}$ for $\rho_{\rm b}=180$, 
magenta squares indicate the estimated value for $\eta^*$.
(b) $\eta^*$ for different $\rho_{\rm b}$ as a function of the number of particles in the blob $N_{\rm b}$.}
\label{fig3}
\end{figure}

Reversals due to the presence of an obstacle become exceedingly rare when decreasing $\eta$. 
To probe whether they can be triggered at low noise strengths,
we study the fate of polar flocks in which we artificially introduce a dense blob of particles
orientated against the main flow, and watch the subsequent evolution (details on our procedure can be found in \cite{SUPP}). 
As for the fixed obstacle, we limit ourselves to circular blobs
of diameter $\sigma_{\rm b}$ and density $\rho_{\rm b}$, and work with system sizes much larger than
$\sigma_{\rm b}$ for which the number of particles in the blob remains much smaller than $N$.
In the range of $\eta$ values where we can observe reversals triggered by an obstacle, 
a large-enough and dense-enough blob reverses the initial flow. 
These blob-induced reversals are similar to obstacle-induced ones, the main difference being that the blob vanishes
completely after its introduction. 
As $\eta$ is decreased, though, a given blob does not always lead to a reversal (see Movie~S4 in \cite{SUPP}).
We find that $P_{\rm rev}$, the probability to reverse the flow upon the introduction of a given blob,
varies from 0 to 1 when $\eta$ is increased, in a manner consistent with a hyperbolic tangent  (Fig.~\ref{fig3}(a)). 
This allows to define a transitional value $\eta^*$  (at $P_{\rm rev}=\tfrac{1}{2}$), 
and to see how it varies with $\rho_{\rm b}$ and $\sigma_{\rm b}$. 
As expected, this procedure works not only for the $\eta$ values `accessible' with a fixed obstacle, but
also with smaller values. 
However, as shown in Fig.~\ref{fig3}(b), we identify a minimal value $\eta\simeq0.40$ below which
no blob, however dense,  triggers a reversal. 
This suggests that only a fraction of the Toner-Tu phase is destroyed by a single localized obstacle.

%%%%%%%%%%%%%%%%%%%%%%%%%%%%%%%%%%%%%%%%%%%%%%%%%%%%
%%%%%%%%%%%%%%%%%%%%%%%%%%%%%%%%%%%%%%%%%%%%%%%%%%%%

%%%%%%%%%%%%%%%%%%%%%%%%%%%%%%%%%%%%%%%%%%%%%%%%%%%%
\begin{figure}[t!]
\includegraphics[width=\columnwidth]{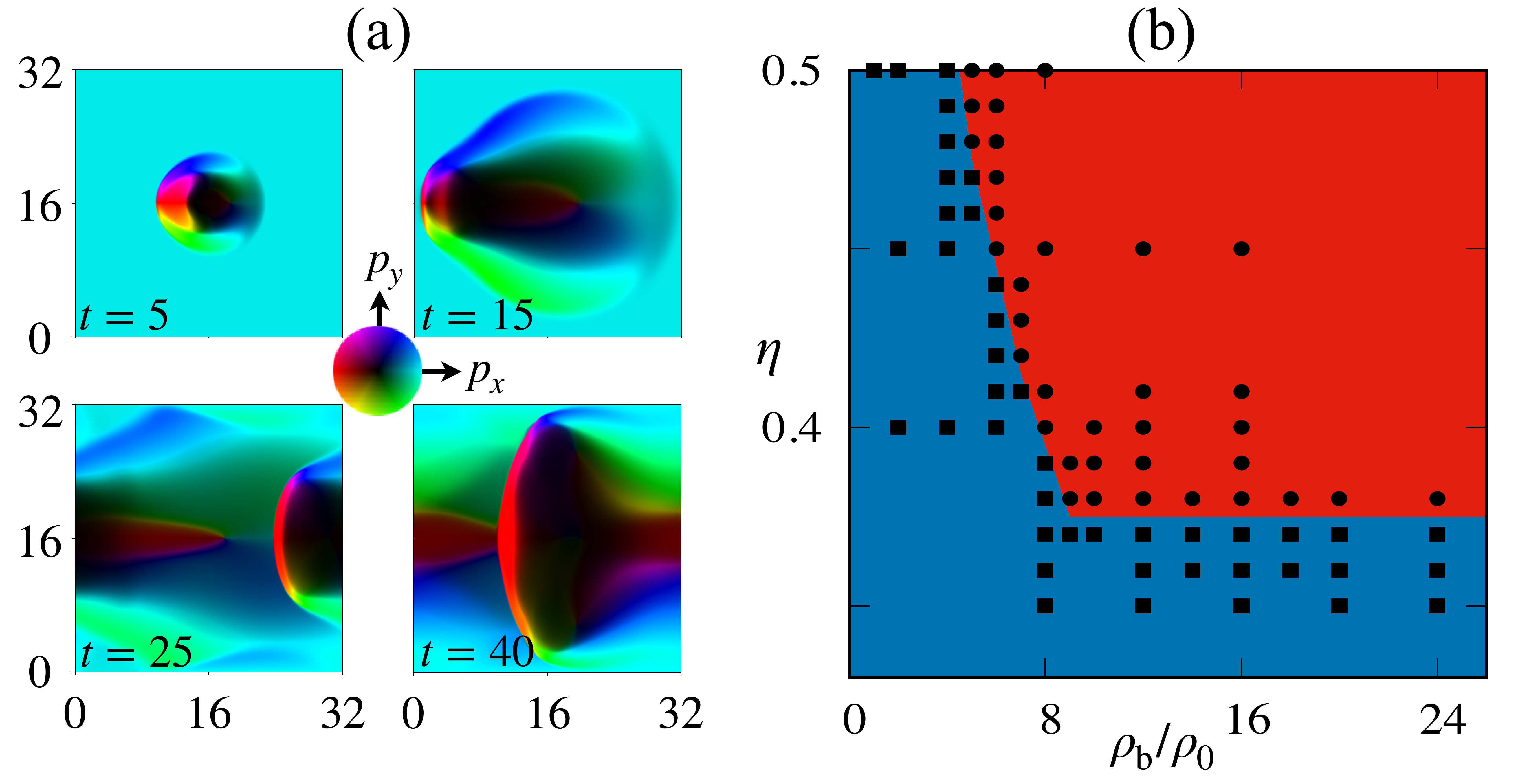}
\caption{Simulations of the Boltzmann equation~\eqref{hierarchy} truncated at $K=4$ at system density $\rho_0 = 2$ and blob radius $\sigma_{\rm b} = 2$.
(a) Snapshots showing the evolution of the momentum field ($f_1 = p_x + i p_y$) at noise $\eta = 0.5$ 
after introduction of a blob with density $\rho_{\rm b} = 8\rho_0$ in the ordered solution at time $t = 0$.
Note that the bivariate color map leads dilute disordered regions to appear darker.
(b) Fate of blob in the $(\rho_{\rm b},\eta)$ plane:
dots mark parameters leading to a reversal, while squares indicate where no reversal occurs.
More details about the numerical protocol in~\cite{SUPP}.
}
\label{fig4}
\end{figure}
%%%%%%%%%%%%%%%%%%%%%%%%%%%%%%%%%%%%%%%%%%%%%%%%%%%%

We finally explore whether continuous kinetic or hydrodynamic theories of polar flocks 
can also account for the phenomena reported above. For numerical convenience, we did not
consider a fixed obstacle but only studied the fate
of an initial small circular blob oriented against the main order. 
Remarkably, when using standard hydrodynamic Toner-Tu equations \cite{SUPP}
no blob, however dense, seems able to trigger a growing counter-propagating band (and thus a reversal) anywhere
in the ordered phase. The initial blob splits into two wings which `diffuse away' (not shown).
On the other hand, reversals do occur when considering low-level truncations of the 
Boltzmann equation for the Vicsek/polar class. 
This equation has been introduced elsewhere and mostly used to derive hydrodynamic theories 
\cite{bertin2006boltzmann,bertin2009hydrodynamic,peshkov2014boltzmann,DADAM_LesHouches}. 
Here we only provide a sketch, with all details given in \cite{SUPP}.
The Boltzmann equation governs the one-body probability distribution function $f({\bf r},\theta,t)$ of finding a particle with velocity orientation $\theta$ at position ${\bf r}$ and time $t$:
\begin{equation}
\partial_t f + v_0 \,{\bf e}(\theta) \!\cdot\! \nabla f  = 
 I_{\rm sd}[f] + I_{\rm co}[f]
\end{equation}
where ${\bf e}(\theta)$ is the unit vector along $\theta$, and $I_{\rm sd}$ and $I_{\rm co}$ are self-diffusion and collision integrals. 
Expanding $f$ in angular Fourier modes $f({\bf r},\theta,t) = \tfrac{1}{2\pi} \sum_k f_k({\bf r},t)\exp(-i k \theta)$, the Boltzmann equation is turned into a hierarchy of coupled partial differential equations 
for the complex fields $f_k$:
\begin{equation}
 \partial_t f_k + \tfrac{1}{2} \left( \triangledown^*\! f_{k+1} \!+\! \triangledown f_{k-1} \right)  =
 \alpha_k f_k  + \sum_q J_{k,q}  f_q f_{k-q}
 \label{hierarchy}
\end{equation}
where $\triangledown=\partial_x +i\partial_y$, and $\alpha_k$ and $J_{k,q}$ are coefficients depending on parameters given in \cite{SUPP}.
The Toner-Tu hydrodynamic theory is recovered when truncating and closing \eqref{hierarchy} using a scaling ansatz
which essentially sets all $f_{k>2}=0$, and enslaves $f_2$, the nematic order field, to the polar field $f_1$.
Here, after realizing that this theory cannot account for reversals, we considered abrupt truncations 
of \eqref{hierarchy} at order $K$, simply setting $f_{k>K}=0$ and integrating the remaining set of $K+1$ equations
(for numerical details, see \cite{SUPP}). 

The lowest order $K$ at which we could observe a full reversal is $K=4$ 
\footnote{We also observed reversals using higher-order truncations. But the necessary numerical resolution
is quickly prohibitively high.}. 
Fig.~\ref{fig4}(a) shows snapshots of the evolution of an initial blob in this case 
(see also Movie~S5 in \cite{SUPP}), which is remarkably similar to that observed with the Vicsek model.
That our set of equations is able to reproduce the development of a reversal is testimony 
that this dynamics is essentially deterministic. 
This allows us to scan parameter space systematically using single runs. 
In Fig.~\ref{fig4}(b) we show the result of such an exploration in the $(\eta,\rho_{\rm b})$ plane.
We find an abrupt limit in $\eta$ below which no blob can lead to a reversal (see Movie~S6 in \cite{SUPP}), 
again in agreement with the microscopic-level results shown in Fig.~\ref{fig3}(b).

To summarize, we have shown that arbitrarily large polar flocks are 
susceptible to the presence of a
single small obstacle: In a fairly wide region of the ordered Toner-Tu phase, the obstacle 
triggers counter-propagating dense bands leading to reversals of the flow.
We observed reversals at the microscopic level using 
 two distinct rules for particle-obstacle collisions, and also 
in truncations of the relevant Boltzmann equation. This stresses the genericity of reversals.
However, we were unable to observe them within the standard Toner-Tu hydrodynamic theory.
This may be a manifestation of the limitations of simple hydrodynamic theories to account for 
highly nonlinear spatiotemporal dynamics (see \cite{cai2019dynamical} for a similar observation in the context of active nematics).

In very large systems we observe complex chaotic dynamics where the bands 
triggered initially interact among themselves and with the obstacle. Increasing system size,
the homogeneous Toner-Tu fluid is never recovered.
Extrapolating to the infinite-size limit, the global order is thus broken. 
As a matter of fact, one can wonder whether the complicated band chaos observed 
could be sustained even in the absence of the obstacle. 
This important question, the answer to which looks so far only accessible via the observation of extremely
large systems, is left for future studies.

\acknowledgments
We thank Yu Duan and Yongfeng Zhao for useful comments. 
We acknowledge generous allocations of cpu time on the Living  Matter  Department  cluster  in  MPIDS,  and on  Beijing CSRC’s Tianhe supercomputer. 
The work was supported by the EU’s Horizon 2020 Program (Grant FET-OPEN 766972-NANOPHLOW to I.P and J.D.),
the National Natural Science Foundation of China (Grants No. 11635002 to
X.-q.S. and H.C., No. 11922506 and No. 11674236 to X.-q.S., No. 11874398 and 12034019 to J.D.),
the Strategic Priority Research Program of the Chinese Academy of Sciences (Grant XDB33000000 to J. D.), 
and an international collaboration grant from the K. C.Wong Education Foundation (to J.D.).
I.P. acknowledges support from Ministerio de Ciencia, Innovaci\'on y Universidades MCIU/AEI/FEDER 
(Grant No. PGC2018-098373-B-100 AEI/FEDER-EU), Generalitat de Catalunya (project No. 2017SGR-884), 
and the Swiss National Science Foundation (Project No. 200021-175719).

\bibliography{./Biblio-current.bib}

\end{document}

\title{Small Obstacle in a Large Polar Flock \\ -- Supplementary Material--}

\author{Joan Codina}
\affiliation{Wenzhou Institute, University of Chinese Academy of Sciences, Wenzhou 325001, China}
\affiliation{Key Laboratory of Soft Matter Physics, Institute of Physics, Chinese Academy of Sciences, Beijing 100190, China}

\author{Beno\^{\i}t Mahault}
\affiliation{Max Planck Institute for Dynamics and Self-Organization (MPIDS), 37077 G\"ottingen, Germany}

\author{Hugues Chat\'{e}}
\affiliation{Service de Physique de l'Etat Condens\'e, CEA, CNRS Universit\'e Paris-Saclay, CEA-Saclay, 91191 Gif-sur-Yvette, France}
\affiliation{Computational Science Research Center, Beijing 100193, China}
\email{hugues.chate@cea.fr}

\author{Jure~Dobnikar}
\affiliation{Key Laboratory of Soft Matter Physics, Institute of Physics, Chinese Academy of Sciences, Beijing 100190, China}
\affiliation{School of Physical Sciences, University of Chinese Academy of Sciences, Beijing 100049, China}
\affiliation{Songshan Lake Materials Laboratory, Dongguan, Guangdong 523808, China}
\email{jd489@cam.ac.uk}

\author{Ignacio Pagonabarraga}
\affiliation{Departament de F\'{\i}sica de la Mat\`eria Condensada, Universitat de Barcelona, 08028 Barcelona, Spain}
\affiliation{Universitat de Barcelona Institute of Complex Systems,
%Universitat de Barcelona, 
08028 Barcelona, Spain}
\affiliation{Centre Europ\'{e}en de Calcul Atomique et Mol\'{e}culaire, Ecole Polytechnique F\'{e}d\'{e}rale de Lausanne, 1015 Lausanne, Switzerland}
\email{ipagonabarraga@ub.edu}

\author{Xia-qing Shi}
\affiliation{Center for Soft Condensed Matter Physics and Interdisciplinary Research, Soochow University, Suzhou 215006, China}
\email{sxiaqing@foxmail.com}

\date{\today}

\maketitle

\section*{Modeling and numerical details}

\begin{figure*}[h]
	\centering
	\includegraphics*[width=\textwidth]{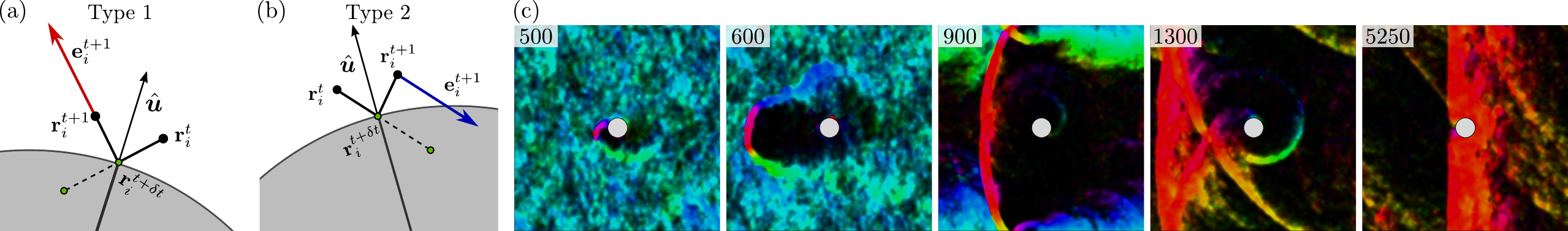}
	\caption{\label{fig:fig1} Sketch of the collision rules. A particle at ${\bf r}_i^{t}$ that would be streamed inside the obstacle is reflected when it contacts the surface of the obstacle at ${\bf r}^\star = {\bf r}_i^{t+\delta t}$, and travels to ${\bf r}_i^{t+1}$. The resulting orientation of the Vicsek particle follows the direction of the reflected trajectory for Type 1 (panel (a)), or the orientation before the collision for Type 2 (panel (b)).
(c) snapshots taken during a reversal as in Fig.~1d of the main text, but with Type~1 collision rules with the obstacle.}
\end{figure*}

\noindent{\bf Collision details.}
We define a collision between a Vicsek particle at position ${\bf r}_i^t$ at time $t$ and an obstacle with center at ${\bf X}$ and radius $\sigma/2$ if the next calculated position of the particle ${\bf r}_i^{t+1} = {\bf r}_i^{t+1} + v_0 {\bf e}_i^{t+1}$ lies inside the obstacle, {\it i.e.} if $\|{\bf X} - {\bf r}_i^{t+1}\|^2 < (\sigma/2)^2$.
When this happens, we locate  ${\bf r}_i^{t+\delta t}$ the point of collision on the boundary of the obstacle,
and the fraction of the time step $\delta t$ required to advance from ${\bf r}_i^t$ to ${\bf r}_i^{t+\delta t}={\bf r}_i^t + v_0 \delta t{\bf e}_i^{t+1}$:
\begin{equation}
v_0	\delta t = -\left({\bf r}^t_i - {\bf X}\right)\cdot {\bf e}_i^{t+1} - \left[\left(\left({\bf r}^t_i - {\bf X}\right)\cdot {\bf e}_i^{t+1} \right)^2+ \left({\bf r}^t_i - {\bf X}\right)^2-\left(\sigma/2\right)^2\right]^{1/2}
\end{equation}
From the collision point we compute the vector normal to the surface of the obstacle $\hat{\bm u} = \vartheta({\bf X} - {\bf r_i}^{t+\delta t})$.
We then reverse the component of the  particle normal to the obstacle surface ${\bf v}_i = {\bf e}_i^{t+1} - 2 \left({\bf e}_i^{t+1}\cdot \hat{\bm u}\right)\hat{\bm u}$. 
The particle's position is then streamed along ${\bf v}$ from the collision point ${\bf x}_i^{t+\delta t}$
during the remaining fraction of the time step 
$(1-\delta t)$: ${\bf r}_i^{t+1} = {\bf r}_i^{t+\delta t} + v_0 (1-\delta t){\bf v}_i$. 
Finally, at the end of the collision process we update the orientation of the particle either in the direction of the reflected velocity, ${\bf e}_i^{t+1}\rightarrow {\bf v}_i$ (Type 1), 
or keep the orientation ${\bf e}_i^{t+1}$ (Type 2).  
Figure~\ref{fig:fig1}(a,b) provides a sketch of the the collision rules.

As mentioned in the main text, even though Type 1 and Type 2 collision rules may appear very different, they yield 
essentially the same phenomenology. In Fig.~\ref{fig:fig1}(c), we show a series of snapshots recorded during
a reversal observed in a system similar to that shown in Fig.~1 of the main text, but with Type 1 collisions between
particles and the obstacle.

%%%%%%%%%%%%%%%%%%%%%%%%%%%%%%%
%%%%%%%%%%%%%%%%%%%%%%%%%%%%%%%

\noindent{\bf Blob reversals.}
In a homogeneous order steady-state with global polarization $\bm \Psi$ we insert $N_{\rm b}$ particles 
with orientation $-\hat{\bm \Psi}$ at random locations within a circle of diameter $\sigma_{\rm b}$. 
The fate of the blob leads to either a complete reversal of the global polarization or the fading of the dense front back into the fluid with no major change in the global polarization, see Movie~S4. 
%%%XXX show movies in each case?

%%%%%%%%%%%%%%%%%%%%%%%%%%%%%%%
\begin{figure}[h]
	\includegraphics[width=\columnwidth]{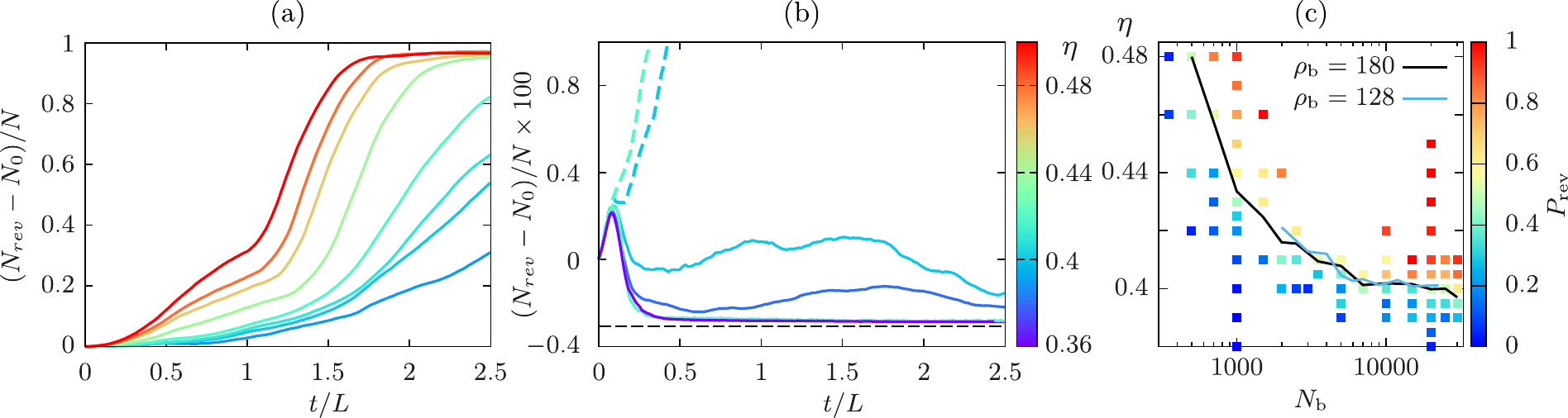}
	\caption{\label{fig:fig2}
(a,b)	Evolution of $(N_{\rm rev}(t)-N_0)/N$, the normalized fraction  of reversed particles after insertion of a blob, where $N_0$ is the number of particles initially present in the blob area that were already oriented against the main flow (i.e. those such that $\Psi_0 \cdot {\bf e}_i^t <0$ at the time of introduction of the $N_{\rm b}$ blob particles). 
(a): blobs that led to complete reversals. 
(b): blobs that did not lead to a reversal (solid lines), together with the trace of 2 successful blobs (dashed lines).
(c) Reversal probability $P_{\rm rev}$ for $\rho_{\rm b}=180$, and the estimated line $P_{\rm rev}=1/2$ in black. In a solid blue line the estimated $P_{\rm rev}=1/2$ line for $\rho_{\rm b}=128$.}
\end{figure}
%%%XXX parameters??!!
%%%%%%%%%%%%%%%%%%%%%%%%%%%%%%%

The evolution of the number of reversed particles at a given time after the blob insertion determines whether the system reverses, or not. Given the orientation at the insertion time, ${\bm \Psi}_0$, we define the number of reversed particles $N_{\rm rev}(t)$ counting the particles that travel against ${\bm \Psi}_0$ at time $t$, $N_{\rm rev}(t) = \sum_i H\left({- \bm \Psi_0 \cdot {\bf e}^{t}_i}\right)$, with $H(x)$ the Heaviside step function.

In Fig.~\ref{fig:fig2} we show the evolution of $N_{\rm rev}$, and distinguish between the blobs that lead to a global reversal $N_{\rm rev}\rightarrow N$, and those who do not.
In this latter case, $N_{\rm rev}$ initially increases and rapidly dies out. 

We exploit these curves to determine a criterion to estimate whether a blob will lead, or not, to full reversal 
without having to simulate the whole process.
Since we are interested in small blobs, $N_{\rm b}/N < 10^{-2}$, we stop the simulation either if $N/N_{\rm rev}>0.2$ (reversal), or if $(N_{\rm rev}-N_0)<0$ (non-reversal).

The insertion of blobs into the homogeneous fluid at different noise strengths $\eta$ permits us to estimate the fraction of blobs that lead to a reversal, $P_{\rm rev}$, for any given $\rho_{\rm b}$, $N_{\rm b}$ pair. In Figure \ref{fig:fig2}(c) we display the measured $P_{\rm rev}$ for $\rho_{\rm b}=180$ and estimate the isoline $P_{\rm rev}=1/2$ that defines the critical value $\eta^\star$ for different blob sizes.

\section*{Boltzmann equation and hydrodynamic theory for polar flocks}

In this Section we give additional details on the kinetic and hydrodynamic descriptions of polar active matter achieved via the Boltzmann-Ginzburg-Landau (BGL) approach.

\vspace{12pt}
\noindent{\bf The Boltzmann equation.} 
As in the main text, we denote $f(\bm r,\theta,t)$ the probability density for a Vicsek-style particle to be at position $\bm r$ with self-propulsion orientation $\theta$ at time $t$.
Following~\cite{peshkov2014boltzmann,DADAM_LesHouches}, this distribution evolves according to the Boltzmann equation
\begin{equation}
 \partial_t f({\bm r},\theta,t) = -v_0 {\bm e}(\theta)\cdot \nabla f({\bm r},\theta,t) + I_{{\rm sd}}[f]+I_{{\rm co}}[f]\,,
	\label{eq_Boltzmann}
\end{equation} 
whose terms on the rhs account, from left to right, respectively for self propulsion with constant speed $v_0$, 
tumblings at rate $\lambda$ and angular noise distribution $P(\theta)$, and binary collisions resulting in polar alignment of interacting particles.
The self-diffusion and collisional integrals read
\begin{align*}
I_{\rm sd}[f] & = \lambda\int_0^{2\pi} d\theta' f({\bm r},\theta',t) [ P(\theta - \theta') - \delta(\theta - \theta')] \,, \\
I_{\rm col}[f] & = 2r_0v_0 \int_{0}^{2\pi} d\theta_1 \int_{0}^{2\pi} d\theta_2 K(\theta_2 \!-\! \theta_1)f({\bm r},\theta_1,t) f({\bm r},\theta_2,t) 
\left( P\left[\theta - \Psi(\theta_1,\theta_2)\right] - \delta(\theta - \theta_1) \right) ,  \label{Icoll}
\end{align*}
where $r_0$ is the interaction radius, 
$K(\Delta) \equiv 2|\sin(\Delta/2)|$ and $\Psi(\theta_1,\theta_2) = {\rm Arg}[e^{i\theta_1} + e^{i\theta_2}]$ respectively stand for the collision kernel and alignment rule. 
For convenience we rescale space, time, and $f$ as
\begin{equation*}
t \to \lambda^{-1} t , \quad \nabla \to \lambda v_0^{-1} \nabla, \quad f \to \rho_0 f ,
\end{equation*}
with $\rho_0$ the mean particle density.
Such rescaling leads Eq.~\eqref{eq_Boltzmann} to depend on only two dimensionless parameters: the variance of the angular noise distribution $\eta^2$, and
$\tilde{\rho}_0 \equiv 2 r_0 v_0 \rho_0 / \lambda$ which we will denote as $\rho_0$ in the following for simplicity and since it plays the role of a nondimensional particle density.

In angular Fourier space, i.e. considering $f({\bm r},\theta,t) = \tfrac{1}{2\pi}\sum_k f_k(\bm r,t) e^{-i k \theta}$, 
Eq.~\eqref{eq_Boltzmann} turns into the hierarchy for the modes  given in Eq.~(3) of the main text, with $\alpha_k = P_k - 1$, $J_{k,q} = \rho_0(P_k I_{k,q} - I_{0,q})$ and
\begin{equation*}
I_{k,q} = \begin{cases}  \frac{4}{\pi}\frac{ 1 - (k-2q)(-1)^{q}\sin\left(\frac{k\pi}{2}\right)}{ 1 - (k-2q)^2} & (\vert k-2q \vert \ne 1) \\
\frac{2}{\pi} & {\rm otherwise} \end{cases} ,
\end{equation*}
and where $P_k$ denotes the $k^{\rm th}$ Fourier mode of the noise distribution $P(\theta)$. 
For all simulations, $P$ was chosen to be a wrapped normal noise distribution such that $P_k = \exp(- k^2 \eta^2 / 2)$.

\vspace{12pt}
\noindent{\bf Numerical simulations of the Boltzmann equation.}  
A straightforward way to numerically integrate the Boltzmann hierarchy is to truncate it by keeping the modes up to a certain order $K$.
(Note that since $f$ is real the modes satisfy $f_{-k} = f_{k}^*$.)~\cite{mahault2018PhD}.
Setting all $f_{k > K} = 0$, we therefore end up with $K+1$ coupled nonlinear partial differential equations for complex fields.
This system of equations was integrated via a pseudo-spectral method --linear (resp. nonlinear) terms were computed in spatial Fourier (resp. real) space--
while time stepping was performed via an explicit $4^{\rm th}$ order Runge-Kutta method.
Due to the steep gradients created by the introduction of the blob, high spatial and temporal resolutions were required to stabilize 
the simulations and resolve correctly the counter-propagating band fronts. 
Typically, we used $dx = \tfrac{1}{32}$ and $dt = 10^{-4}$ and checked that the results presented in the main text remain stable upon 
changing these values within reasonable range.
Despite the high resolution of simulations, blob initial conditions could lead to unphysical solutions (typically with negative density $f_0$) at high enough blob densities.
As regularizing these solutions would have required prohibitively large spatial and temporal resolutions, 
we instead enforced $f_0 \ge 0$ by setting $f_0(\bm r,t)$ to zero when it reaches negative values, 
while the mean density $\langle f_0 \rangle_{\bm r} = 1$ was kept constant by redistributing the mass difference over all other space points.
As this correction scheme is implemented at each simulation steps, 
for $dt$ small enough it concerns only a few points and the total redistributed mass remains small compared to 1.
We moreover checked that for moderate values of blob density where simulations are stable 
although negative values of $f_0$ might occur during transients, the behavior of the system was similar with and without enforcing $f_0 \ge 0$.

\vspace{12pt}
\noindent{\bf Initialization of a blob.}
To simulate the introduction of a circular blob of uniform density $\rho_{\rm b}$ at the center of a domain of physical dimensions $L_x \times L_y$ 
discretized over a grid of dimensions $N_x \times N_y$, we first define 
${\cal S}_{\rm b} \equiv \{ (x_i,y_i), \sqrt{ (x_i - L_x/2)^2 + (y_i - L_y/2)^2} < \sigma_{\rm b} \}_{i = 1,\ldots,N_x N_y}$
as the set of the square grid points belonging to the blob of radius $\sigma_{\rm b}$.
From total mass conservation, the density of the surrounding Toner Tu liquid is given by
\begin{equation*}
\rho_{\rm TT} = \frac{\rho_0 N_x N_y - \rho_{\rm b} n({\cal S}_{\rm b})}{N_x N_y - n({\cal S}_{\rm b})},
\end{equation*}
with $n({\cal S}_{\rm b})$ the number of points in ${\cal S}_{\rm b}$.
The initial condition is thus built evaluating the homogeneous ordered solutions $\bar{f}_k^{\rm TT}$ and $\bar{f}_k^{\rm b}$ of the Boltzmann hierarchy for all $k \le K$
at densities respectively $\rho_{\rm TT}$ and $\rho_{\rm b}$ and with opposite global polarity orientations. We then assign
\begin{equation*}
f_k(\bm r_i,t=0) = \begin{cases} \bar{f}_k^{\rm b} & \bm r_i \in {\cal S}_{\rm b} \\ \bar{f}_k^{\rm TT} & {\rm otherwise}  \end{cases} .
\end{equation*}
Note that the above protocol can be adapted to consider less abrupt initial conditions, e.g. by imposing a smooth initial blob density profile.
However, this does not lead to significant changes in the short time dynamics following the blob introduction.

\vspace{12pt}
\noindent{\bf Choice of parameters.}
For simplicity we fixed the mean rescaled density $\rho_0 = 2$ and the bob radius $\sigma_{\rm b} = 2$, which leaves two control parameters: 
the angular noise strength $\eta$ and the blob density $\rho_{\rm b}$.
For any value of $K$, the phase diagram of the Boltzmann equation~\eqref{eq_Boltzmann} in the moderate noise regime always shows the three phases
observed in simulations of the Vicsek model:
disordered gas and ordered liquid respectively at large and low noises, while the two are separated by the phase coexistence phase (Vicsek bands).
However, for small $K$ truncations the precise location of the band phase may vary with $K$, 
so that in order to ensure that simulations were done in a stable homogeneous Toner Tu liquid we evaluated its linear stability following the procedure presented in~\cite{mahault2018PhD}.
Values of $\eta$ were then chosen to remain far enough from the onset of instability. 

For $K=1$ and $3$, due to the absence of stabilizing nonlinearities the homogeneous solutions of the Boltzmann equation are not bounded, so that these pathological cases were discarded.
Considering $K=2$ --so three coupled equations for the density, polar and nematic fields-- we found that the numerical protocol defined above did not lead to any reversal for a broad range of $\eta$ and $\rho_{\rm b}$ values. 
Indeed, at this order the initial band front created by the blob only survives a finite time and never manages to span the whole system.
On the other hand, considering $K=4$ we could easily observe reversals for $\rho_{\rm b} > 4 \rho_0$ over a certain range of $\eta$ values (Fig. 4 of the main text).
Reversals are also found at larger values of $K$ and we checked that $K = 6$ in particular leads to similar results as $K=4$.
However, as increasing $K$ goes hand in hand with increasing resolution, $K > 6$ are hardly accessible numerically. 
Nevertheless, for $\rho_0 = 2$ and $\eta \ge 0.3$ the linear stability analysis of the homogeneous ordered solution of the Boltzmann equation 
essentially give identical results for all $K \ge 4$~\cite{mahault2018PhD}, so that we do not expect  significant variations of the results presented in Fig. 4 for larger $K$.

\vspace{12pt}
\noindent{\bf Blob-induced reversals.}
Reversal events in Fig. 4(b) of the main text were defined by visual inspection of the initial counter-propagating front created by the blob.
In absence of reversals the front quickly breaks, usually long before it manages to span the system and loop back to its initial position.
We therefore defined a reversal event as when a counter-propagating front survives upon crossing a full system size.
We found that this criterion leads to results robust with varying system size, as well as the resolution,
even for large $\rho_{\rm b}$ and $\eta$ close to the limit value $\simeq 0.37$. 

\vspace{12pt}
\noindent{\bf The hydrodynamic level.}
The Boltzmann equation~\eqref{eq_Boltzmann} is usually used as an intermediate step to derive the simpler hydrodynamic description of dry active matter~\cite{peshkov2014boltzmann,DADAM_LesHouches}.
At the onset or polar order $(|f_1| \sim \varepsilon\ll 1)$, it can indeed be shown that the Fourier modes, spatial and temporal derivatives satisfy the following scaling ansatz
\begin{equation*}
|f_0 - 1| \sim \varepsilon, \quad |f_k| \sim \varepsilon^{|k|} \; (k > 0), \quad \partial_t \sim \partial_{x,y} \sim \varepsilon ,
\end{equation*}
such that keeping terms only up to order $\varepsilon^3$ leads to retain a set of closed equations for $f_{0,1,2}$. 
The nematic field $f_2$ can moreover be shown to be fast and is enslaved to the slow modes of the dynamics: $f_0$ and $f_1$.
The resulting equations are given in complex notations by
\begin{subequations}
\label{eq_hydro_polar}
\begin{align}
 \partial_t f_0 & = -\Re\left(\triangledown^* f_1\right) \,, \\
 \partial_t f_1 & = -\frac{1}{2}\triangledown f_0 + \left(\mu_1[f_0] - \xi |f_1|^2\right) f_1 + \nu \Delta f_1 
- \kappa_1 f_1 \triangledown^* f_1 - \kappa_2 f_1^* \triangledown f_1  \,, 
\end{align}
\end{subequations}
where the coefficients read
\begin{align*}
\mu_1[\rho] = & P_1 - 1 + \frac{4}{\pi}\left(P_1 - \frac{2}{3}\right)\rho_0\rho \,, &
\mu_2 = &P_2 - 1 - \frac{8}{15\pi}(7 + 5P_2)\rho_0 \,, \\
\xi = & -\frac{16(5P_1-2)(3P_2+1)\rho_0^2}{15\pi^2 \mu_2} \,, &
\nu = &- \frac{1}{4\mu_2} \,, \\
\kappa_1 = &-\frac{4(1+3P_2)\rho_0}{3\pi \mu_2} \,, &
\kappa_2 = &\frac{2(5P_1-2)\rho_0}{5\pi \mu_2} \,.
 \end{align*}
Eqs.~\eqref{eq_hydro_polar} were numerically integrated using a similar method as described above for the truncated Boltzmann hierarchy. 
Consistently with the Boltzmann results obtained at $K=2$, we find that at the hydrodynamic level an initial blob never triggers a counter propagating front strong enough to revert the direction of the global order. 
We moreover checked that this feature persists at the stochastic hydrodynamic level where the equation for $f_1$ 
is endowed with a multiplicative noise term of variance $\propto f_0$.

\section*{Description of movies}

\noindent
{\bf Movie S1:} Temporally isolated reversal nucleated from the homogeneous fluid (Vicsek model in a periodic square domain, with a circular central obstacle, $L=1024$, $\eta=0.5$, $\sigma=22$, momentum field represented as in Fig.~1 of the main text)

\vspace{12pt}
\noindent
{\bf Movie S2:} Same as Movie 1, but showing multiple reversals, including some nucleated from a passing band. 

\vspace{12pt}
\noindent
{\bf Movie S3:} Long transient following an initial reversal in a large system
(Vicsek model in a periodic square domain, with a circular central obstacle, $L=4096$, $\eta=0.5$, $\sigma=35$ [Parameters of Figure 2c of main text]).

\vspace{12pt}
\noindent
{\bf Movie S4:} Evolution an initial blob for different values of $\eta$
(Vicsek model in a periodic square domain, 
$L=1024$, $\rho_b=180$, $N_{b} = 20000$)

\vspace{12pt}
\noindent
{\bf Movie S5:} Reversal from a blob in simulations of the Boltzmann equation truncated at order $K = 4$. Parameters are $\eta = 0.5$ and $\rho_{\rm b} = 8$. 

\vspace{12pt}
\noindent
{\bf Movie S6:} Same as Movie S5 but showing the absence of reversal at $\eta = 0.35$.

\bibliography{./Biblio-current.bib}